\documentclass[aps,pra,reprint,preprintnumbers,showpacs,showkeys,superscriptaddress]{revtex4-1}
\usepackage{amsmath}
\usepackage{amssymb}
\usepackage{bm}
\usepackage{graphicx}
\usepackage{subfigure}

\begin{document}

\title{Spontaneous breaking of continuous translational invariance}

\author{Haruki Watanabe}
\email{hwatanabe@berkeley.edu}
\affiliation{Department of Physics, University of Tokyo, Hongo, Tokyo 113-0033, Japan}
\affiliation{Department of Physics, University of California, Berkeley, California 94720, USA}

\author{Tom\'{a}\v{s} Brauner}
\email{tbrauner@physik.uni-bielefeld.de}
\affiliation{Faculty of Physics, University of Bielefeld, 33615 Bielefeld, Germany} 
\affiliation{Department of Theoretical Physics, Nuclear Physics Institute ASCR, 25068 \v Re\v z, Czech Republic}

\begin{abstract}
Unbroken continuous translational invariance is often taken as a basic assumption in discussions of spontaneous symmetry breaking (SSB), which singles out SSB of translational invariance itself as an exceptional case. We present a framework which allows us to treat translational invariance on the same footing as other symmetries. It is shown that existing theorems on SSB can be straightforwardly extended to this general case. As a concrete application, we analyze the Nambu--Goldstone modes in a (ferromagnetic) supersolid. We prove on the ground of the general theorems that the Bogoliubov mode stemming from a spontaneously broken internal U$(1)$ symmetry and the longitudinal phonon due to a crystalline order are distinct physical modes.
\end{abstract}

\preprint{BI-TP 2011/49}
\maketitle


\section{Introduction}

The low-energy physics of systems with spontaneous symmetry breaking (SSB) is dominated by the associated soft modes, the Nambu--Goldstone (NG) bosons. The existence of at least one such a soft mode is guaranteed by the celebrated Goldstone theorem~\cite{Goldstone:1961eq,*Goldstone:1962es} (see also Ref.~\cite{Guralnik:1968gu} for a comprehensive early review). The question of how many NG bosons there are given the symmetry-breaking pattern is easy to answer in Lorentz-invariant systems. In the general case, it is however difficult; several partial solutions to the problem have been given over the years~\cite{Nielsen:1975hm,Schafer:2001bq,Leutwyler:1993gf} (see also Ref.~\cite{Brauner:2010wm} for a recent review).

What is common to the general results existing in literature is the separate treatment of internal and spacetime symmetries. In particular, it is usually assumed that the ground state of the system has a continuous translational invariance. SSB of translational invariance itself is then treated as an exceptional case (an analysis based on effective field theory was given in Ref.~\cite{Leutwyler:1997hp}). Since only a discrete translational symmetry is needed to have well-defined low-momentum quasiparticles, and thus to discuss the NG bosons, such separation seems to be merely a matter of convenience. 

In the recent paper~\cite{WatanabeBrauner:2011}, we argued that the distinction between internal and spacetime symmetries is somewhat artificial, at least as long as SSB is considered. Instead, we pointed out that the requirement necessary for standard arguments concerning SSB to apply is that the symmetry be \emph{uniform}. This, roughly speaking, means that the infinitesimal symmetry transformation does not depend explicitly on the spacetime coordinates. Formally, it can be encoded in the condition on the translational property of the Noether charge density, Eq.~\eqref{eq:uniform}.

The objective of the present paper is to complement our previous study by a more detailed discussion of spontaneous breaking of translational invariance. Since translational invariance is the only uniform spacetime symmetry, this completes our analysis and shows that all uniform symmetries can be treated on the same footing. At the same time, it provides a generalization of the other existing results. What we do assume is an unbroken \emph{discrete} translational invariance of the ground state though. This is a minimal necessary requirement for the notion of NG bosons as soft quasiparticles to be meaningful. One can expect that even in the complete absence of translational invariance, SSB will give rise to low-lying excited states in the spectrum; such a generalization, however, goes beyond the scope of this paper.

We would hasten to emphasize that the issue of spontaneous breaking of translational invariance is, of course, not of merely academic interest. Systems where an inhomogeneous state develops spontaneously have been extensively discussed in the literature. A classic example is the inhomogeneous pairing in superconductors in external magnetic field, predicted by Larkin, Ovchinnikov, Fulde, and Ferrell (LOFF)~\cite{Fulde:1964zz,*Larkin:1964zz}. Similar type of pairing can occur in many other systems, ranging from ultracold atomic gases to cold dense quark matter, see Ref.~\cite{Casalbuoni:2003wh} for a review and further references.

In LOFF-type superconductors, the order parameter is usually modulated in one dimension. On the contrary, complete breaking of continuous translational invariance is featured by another prominent example, the supersolid~\cite{Andreev:1969,*Chester:supersolid,*Kim:2004nature,*Kim:2004science}. A supersolid is a state which has both a crystalline order (ordinary long-range order) and a superfluid order (off-diagonal long-range order). In the present paper, we will use it as an example to illustrate the general results concerning spontaneous breaking of translational invariance. Other examples of systems exhibiting spontaneous breaking of translational invariance include stripe ordering in spinor Bose--Einstein condensates~\cite{Wang:2010} and numerous systems in high-energy physics such as the charge or chiral density wave~\cite{Overhauser:1978,*Deryagin:1992rw,*Shuster:1999tn} or the quarkyonic chiral spiral~\cite{Kojo:2009ha,*Kojo:2011cn}.

All in all, it is clearly desirable to have a systematic framework that can be used to analyze the physics of NG modes in systems where translational invariance is broken to a (nontrivial) discrete subgroup. Let us remark that all our general arguments, presented in Section~\ref{sec:general}, are actually valid regardless of whether this breaking is spontaneous or explicit. As a byproduct of our interest in spontaneous breaking of translational invariance, we therefore provide a formalism for description of NG bosons of spontaneously broken internal uniform symmetries in systems with a discrete periodic structure such as crystalline solids or optical lattices~\footnote{These have the same translational properties as intrinsically discrete systems defined on a (spatial) lattice. However, since the extension to such systems would require further generalization of the notation, we exclude them for the sake of simplicity from our considerations.}.

The plan of the paper is as follows. In Section~\ref{sec:general}, we discuss in detail the setup and generalize the existing theorems on NG modes, in particular the Goldstone theorem~\cite{Goldstone:1961eq,*Goldstone:1962es}, the theorem of Sch\"afer~\emph{et al.}~\cite{Schafer:2001bq}, and the Nielsen--Chadha theorem~\cite{Nielsen:1975hm}. Readers not interested in general properties of spontaneous symmetry breaking might prefer to skip this section and proceed directly to Section~\ref{sec:supersolid} where we analyze in detail a model for a two-dimensional (ferromagnetic) supersolid. As another nontrivial example, we argue how translational properties of the system get affected by an external uniform magnetic field and discuss phonons in a charged crystal under the magnetic field (Section~\ref{sec:chargedphonon}). In Section~\ref{sec:summary}, we summarize and conclude.


\section{General theorems}
\label{sec:general}

As a first step towards the generalization of the existing results on NG bosons, we carefully formulate our assumptions. This will in particular clarify the scope of the validity of our arguments presented in the later parts of this section.


\subsection{Assumptions}

First of all, we should emphasize that when we speak of breaking translational invariance, we always have in mind \emph{spatial} translations. Continuous time translation invariance, which guarantees energy conservation, will be assumed throughout this paper without further mentioning it. As to the spatial translations, we assume discrete translational invariance in $d'$ directions and continuous one in the remaining $d-d'$ directions. This requirement is imposed on both the Hamiltonian (or Lagrangian) of the theory and its ground state. The translation symmetry group is thus $\mathcal{G}_{\rm D}^{(d')}\times\mathcal{G}_{\rm C}^{(d-d')}$ where $\mathcal{G}_{\rm D}^{(d')}=\{T_{\bm{R}}\equiv e^{i\bm{P}_1\cdot\bm{R}/\hbar}\,|\,\bm{R}=\sum_{i=1}^{d'}n_i\bm{a}_i\,\,(n_i\in \mathbb{Z})\}$ and $\{\bm{a}_i\}_{i=1}^{d'}$ are the primitive lattice vectors. Furthermore, $\mathcal{G}_{\rm C}^{(d-d')}=\{e^{i\bm{P}_2\cdot\Delta\bm{x}/\hbar}\,|\,\Delta\bm{x}\in\mathbb{R}^{d-d'}\}$ where $\bm P$ is the momentum operator and $\bm P_1,\bm P_2$ its components in $\mathbb R^{d'}$ and $\mathbb R^{d-d'}$, respectively. Likewise, the spatial coordinate vector $\bm x$ can be decomposed as $\bm x=(\bm x_1,\bm x_2)$ where $\bm x_1\in\mathbb{R}^{d'}$ and $\bm x_2\in\mathbb{R}^{d-d'}$. Moreover, the vector $\bm x_1$ can be uniquely represented as $\bm x_1=\bar{\bm x}_1+\bm R$ where $\bm R$ is a lattice vector and $\bar{\bm x}_1$ belongs to the Wigner--Seitz unit cell in $\mathbb{R}^{d'}$. We will sometimes for the sake of brevity refer to the set of all spacetime coordinates, $(t,\bm x)$, using the corresponding regular-weight letter, $x$.

Apart from translational invariance, the system may possess other, internal symmetries at the Lagrangian level, which can be spontaneously broken. Following the argument presented in Ref.~\cite{WatanabeBrauner:2011}, we always assume that such symmetries are uniform, that is, their Noether charge density $j^0(x)$ satisfies the following condition~\footnote{In fact, we should further distinguish which of the first $d'$ coordinates are associated with continuous translational invariance that is spontaneously broken, and which with an intrinsically discrete translational symmetry. Since identities such as Eq.~\eqref{eq:uniform} only rely on the commutation relations of the Noether charge(s), the former type of coordinates actually possess a stronger, ``continuous'' translational property. It is good to keep this in mind, it will nevertheless not be essential for the discussion to follow.},
\begin{multline}
j^0(t,\bm x_1,\bm x_2)\\
=e^{i(Ht-\bm P_2\cdot\bm x_2)/\hbar}T^\dag_{\bm R}\,j^0(0,\bar{\bm x}_1,\bm 0)\,T_{\bm R}\,e^{-i(Ht-\bm P_2\cdot\bm x_2)/\hbar}.
\label{eq:uniform}
\end{multline}
We also assume that the charge associated with a spontaneously broken symmetry (whether internal or translational) is time-independent and is given by an integral of a local charge density, $Q=\int d^d\bm{x}\,j^0(t,\bm x)$.

Finally, we assume that there is a complete set $\{|\xi,\bm{k}\rangle\}$ of simultaneous eigenstates of the Hamiltonian ($H|\xi,\bm{k}\rangle=\varepsilon_{\xi}(\bm{k})|\xi,\bm{k}\rangle$), discrete translations in $\mathbb R^{d'}$ ($T_{\bm{R}}|\xi,\bm{k}\rangle=e^{i\bm{k}_1\cdot\bm{R}}|\xi,\bm{k}\rangle$), and the remaining momentum operators ($\bm{P}_2|\xi,\bm{k}\rangle=\hbar\bm{k}_2|\xi,\bm{k}\rangle)$. The normalization of the states is chosen as $\langle\xi,\bm{k}|\xi',\bm{k}'\rangle=(2\pi)^d\delta_{\xi\xi'}\delta^d(\bm{k}-\bm{k}')$ so that the completeness of the basis is expressed by the relation
\begin{equation}
\sum_{\xi}\int_{(\mathrm{FBZ})}\frac{d^d\bm{k}}{(2\pi)^{d}}|\xi,\bm{k}\rangle\langle \xi,\bm{k}|=\openone.
\label{eq:partition}
\end{equation}
The integration over $\bm k_1$ is performed within the first Brillouin zone (FBZ), that is, the Wigner--Seitz cell of the reciprocal space, while the $\bm k_2$ integration is done over the whole $\mathbb R^{d-d'}$ space. The index $\xi$ may include discrete as well as continuous labels such as the band index, which together with the momentum uniquely identify each eigenstate.

It is customary to characterize SSB by the number of spontaneously broken generators $Q_a$, denoted here as $n_{\text{BS}}$. This can be usually determined as the difference of the dimensions of the symmetry groups of the action and the ground state of the theory. However, for the sake of general arguments, it is more convenient to use the following formal definition: $n_{\text{BS}}$ symmetries are said to be spontaneously broken if there is as set of (quasi)local operators $\{\phi_a(x)\}_{a=1}^{n_{\text{BS}}}$ such that the matrix $M$ given by $iM_{ab}\equiv\langle0|[Q_a,\phi_b(0)]|0\rangle$ is nonsingular.


\subsection{Goldstone theorem}

The Goldstone theorem~\cite{Goldstone:1961eq,*Goldstone:1962es} claims that whenever a uniform symmetry is spontaneously broken, there should be \emph{at least} one massless bosonic state in the spectrum of the theory. The proof proceeds essentially by finding the spectral representation of the commutator of the broken charge density $j^0_a$ and the interpolating field $\phi_b$. Here we adapt the standard argument to account for the possibility of discrete translational invariance and obtain the following spectral representation,
\begin{equation}
\begin{split}
iS&_{ab}(\omega,\bm k)\equiv\int d^{d+1}x\,e^{i(\omega t-\bm k\cdot\bm x)}\langle 0|[j^0_a(x), \phi_b(0)]|0\rangle\\
=&2\pi\sum_{\xi}\Bigl[\delta(\omega-\tfrac1\hbar\varepsilon_{\xi}(\bm{k}))\langle 0|\overline{j_a}_{(\bm{k})}|\xi,\bm{k}\rangle\langle\xi,\bm{k}|\phi_b(0)|0\rangle\\
&-\delta(\omega+\tfrac1\hbar\varepsilon_{\xi}(-\bm{k}))\langle 0|\phi_b(0)|\xi,-\bm{k}\rangle\langle \xi,-\bm{k}|\overline{j_a}_{(\bm{k})}|0\rangle\Bigr],
\end{split}
\label{eq:S}
\end{equation}
which holds provided $\bm k_1$ lies in FBZ. This expression is formally identical to one assuming full continuous translational invariance; all effects of only discrete translational symmetry in $\mathbb R^{d'}$ are hidden in the definition of the charge density averaged over the unit cell,
\begin{equation}
\overline{j_a}_{(\bm{k})}\equiv\frac1{\Omega_{\text{uc}}}\int_{\mathrm{uc}}d^{d'}\bar{\bm{x}}_1\,\,j^0_a(0,\bar{\bm{x}}_1,\bm{0})e^{-i\bm{k}_1\cdot\bar{\bm{x}}_1},
\label{eq:ucaverage}
\end{equation}
where $\Omega_{\mathrm{uc}}$ is the volume of the Wigner--Seitz unit cell, over which the $\bm x_1$ integral is performed (this is indicated by the subscript ``uc''). Note that in the limit of $\Omega_{\mathrm{uc}}\rightarrow0$, $\overline{j_a}_{(\bm{k})}\rightarrow j_a^0(0)$ which leads to the usual spectral representation with the full continuous translational invariance.

The proof of Eq.~\eqref{eq:S} is straightforward but somewhat tedious, and we thus refer the reader to Appendix~\ref{app:derivation} for details. Consider now the limit $|\bm{k}|\rightarrow 0$ of the commutator Fourier transformed only in the spatial coordinates,
\begin{align}
\notag
\lim_{|\bm{k}|\rightarrow 0}iS_{ab}(t,\bm{k})&\equiv\lim_{|\bm{k}|\rightarrow 0}\int d^{d}\bm x\,e^{-i\bm{k}\cdot\bm{x}}\langle 0|[j^0_a(x), \phi_b(0)]|0\rangle\\
&=\langle 0|[Q_a, \phi_b(0)]|0\rangle=iM_{ab}.
\end{align}
Using the assumption that the charges $Q_a$ are time-independent (which is most naturally ensured when $j^0_a$ satisfies a continuity equation together with a spatial current $\bm j_a$, and the corresponding surface term in the integration over space vanishes), we deduce that $\displaystyle\lim_{|\bm{k}|\rightarrow 0}iS_{ab}(\omega,\bm k)\propto\delta(\omega)$. Combining this with the spectral representation~\eqref{eq:S}, we find
\begin{equation}
M_{ab}=2\sum_{\xi|\varepsilon_{\xi}(\bm{0})=0}\mathrm{Im}\left[\langle 0|\overline{j_a}|\xi,\bm{0}\rangle\langle \xi,\bm{0}|\phi_b(0)|0\rangle\right],
\label{eq:M}
\end{equation}
where $\overline{j_a}\equiv \overline{j_a}_{(\bm{0})}$.

The condition of SSB ($\det M\neq0$) ensures that the right-hand side of this equation is nonzero for some $a,b$. One might at first think that this is due to the contribution of degenerate ground states obtained by the action of broken symmetry transformations on $|0\rangle$. However, this contribution drops at large volume $\Omega$ asymptotically as $1/\Omega$ since such degenerate ground states only exist for isolated values of momentum (for which $\bm k_2=\bm0$ and $\bm k_1$ is equal to some of the vectors of the reciprocal lattice)~\cite{Guralnik:1968gu,Lange:1965,*Lange:1966}. It then follows that there must be at least one intermediate excited state $|n,\bm{k}\rangle$ such that: (i) $\langle 0|\overline{j_a}_{(\bm{k})}|n,\bm{k}\rangle\neq0$ and $\langle n,\bm{k}|\phi_b(0)|0\rangle\neq0$ for some $a, b$ in the vicinity of $\bm{k}=\bm0$, and (ii) $\varepsilon_{n}(\bm{0})=0$. Such intermediate states are exactly the Nambu--Goldstone modes associated with spontaneous breaking of the charges $Q_a$. Denoting $|n\rangle\equiv |n,\bm{0}\rangle$ and the number of these states as $n_{\text{NG}}$, Eq.~\eqref{eq:M} becomes
\begin{equation}
M_{ab}=2\sum_{n=1}^{n_{\mathrm{NG}}}\,\,\mathrm{Im}\left[\langle 0|\overline{j_a}|n\rangle\langle n|\phi_b(0)|0\rangle\right].
\label{eq:M2}
\end{equation}

Before concluding the discussion of the Goldstone theorem, let us remark that in case there are long-range interactions in the system, the surface term appearing in the space integral of the continuity equation for the Noether charge density and its current need not vanish. The integral charge then depends on time and massive modes can contribute to Eq.~\eqref{eq:M}. This is why the longitudinal mode in a Coulomb-interacting Wigner solid (in three spatial dimensions) acquires an energy gap, equal to the plasma frequency~\cite{Clark:1958,*Brout:1958}.

Finally, note that the above presented proof of the Goldstone theorem relies essentially on the operator formalism and the translational property~\eqref{eq:uniform} of the charge density as well as the time independence of the integral charge. In concrete applications, it is often advantageous to use an alternative proof based on the effective action formalism~\cite{Goldstone:1962es,Weinberg:1996v2}, which provides a direct connection to the symmetries of the corresponding classical Lagrangian system. For the sake of completeness, we sketch a modification of this proof suitable for systems with only discrete translational invariance in Appendix~\ref{app:effaction}.


\subsection{Theorem of Sch\"afer \emph{et al.}}

In the preceding subsection, we carefully formulated the Goldstone theorem as showing the existence of at least one NG boson. The natural question to ask is, of course, whether anything can be said in general about their actual number, $n_{\text{NG}}$. A common lore in textbooks on relativistic field theory is that assuming Lorentz invariance, $n_{\text{NG}}$ equals the number of spontaneously broken generators, $n_{\text{BS}}$. Nevertheless, there are numerous examples of Lorentz-noninvariant (both relativistic many-body as well as intrinsically nonrelativistic) systems where this simple relation does not hold (see Ref.~\cite{Brauner:2010wm} and references therein).

We start our discussion with the result of Sch\"afer \emph{et al.}~\cite{Schafer:2001bq} who formulated the following theorem~\footnote{Strictly speaking, what they proved is that the number of NG modes is \emph{equal to or greater than} the number of broken symmetries. Here we supplement the theorem with the remaining half of the proof.}: Provided that $\langle 0|[Q_a,Q_b]|0\rangle=0$ for all $a,b$, the number of NG modes $n_{\text{NG}}$ is equal to the number of broken symmetries $n_{\text{BS}}$. Here we generalize this theorem by taking into account the possibility of a discrete translational symmetry in $\mathbb R^{d'}$ (whether as a result of spontaneous breaking of continuous translational invariance or of the intrinsic setup of the system). We follow essentially the argument of the review~\cite{Brauner:2010wm}.

Let us first show that $n_{\mathrm{NG}}\geq  n_{\mathrm{BS}}$. Based on the remaining translational symmetry of the ground state, we infer that $\langle 0|[Q_a,Q_b]|0\rangle\propto \langle 0|[Q_a,\overline{j_b}]|0\rangle$ and, additionally, using Eq.~\eqref{eq:M2},
\begin{equation}
\langle 0|[Q_a,\overline{j_b}]|0\rangle=2i\sum_{n=1}^{n_{\mathrm{NG}}}\mathrm{Im}\left[\langle 0|\overline{j_a}|n\rangle\langle n|\overline{j_b}|0\rangle\right].
\label{eq:CR}
\end{equation}
Defining an $n_{\mathrm{BS}}\times n_{\mathrm{NG}}$ matrix $A$ as $A_{an}=\langle 0|\overline{j_a}|n\rangle$, the assumption that $\langle 0|[Q_a,Q_b]|0\rangle=0$ for all $a,b$ together with Eq.~\eqref{eq:CR} implies that $AA^{\dagger}$ is Hermitian as well as real. It can thus be diagonalized by a real orthogonal matrix $O$ such that $A'{A'}^{\dagger}\equiv OAA^{\dagger}O^{T}=\mathrm{diag}(\lambda_1,\dotsc,\lambda_{n_{\mathrm{BS}}})$, assuming without lack of generality $0\leq \lambda_1\leq \dotsb\leq  \lambda_{n_{\mathrm{BS}}}$. Now if the number of NG bosons were smaller than $n_{\text{BS}}$, the rank of $A$, and hence also the rank of $AA^\dagger$, being at most $n_{\text{NG}}$, would have to be smaller than $n_{\text{BS}}$ as well. The smallest eigenvalue $\lambda_1$ would then necessarily vanish. However, since the orthogonal transformation by $O$ can be interpreted as a change of basis of the broken generators, $Q_a'=O_{ab}Q_b$, this would lead to $\langle 0|\overline{j_1}'|n\rangle=\langle n|\overline{j_1}'|0\rangle=0$ for all $n$, in contradiction with the broken symmetry assumption, $\det M\neq0$. (Note that $M'=OM$ so that determinants of $M$ and $M'$ are equal up to a sign.) Thus, the inequality $n_{\mathrm{NG}}\geq  n_{\mathrm{BS}}$ follows.

In order to prove the opposite inequality, $n_{\mathrm{NG}}\leq n_{\mathrm{BS}}$, observe that $A^{\dagger}A$ is positive semidefinite and Hermitian, and can therefore be diagonalized by a unitary matrix $U$ as ${A''}^{\dagger}A''\equiv U^{\dagger}A^{\dagger}AU=\mathrm{diag}(\rho_1,\dotsc, \rho_{n_{\mathrm{NG}}})$, again assuming without lack of generality that $0\leq\rho_1\leq\dotsb\leq\rho_{n_{\text{NG}}}$. The transformation $A\to A''=AU$ corresponds to a redefinition of the basis of states, $|m''\rangle=|n\rangle U_{nm}$, since $A_{am}''=A_{an}U_{nm}=\langle 0|\overline{j_a}|m''\rangle$ (summations over repeated indices are implied). Now if the number of NG bosons were larger than $n_{\text{BS}}$, we would analogously deduce that $\langle 0|\overline{j_a}|m''\rangle=0$ for all $a$ and for $m=1,2,\dotsc,n_{\mathrm{NG}}-n_{\mathrm{BS}}$. However, these zero modes do not couple to any broken currents and thus do not satisfy the very definition of a NG boson. 

Putting both pieces together, we have proven that the number of NG modes is equal to the number of broken symmetries if $\langle 0|[Q_a,Q_b]|0\rangle=0$ for all $a, b$.


\subsection{Nielsen--Chadha theorem}

Nielsen and Chadha~\cite{Nielsen:1975hm} discovered that the number of NG bosons is tightly related to their dispersion relations. Concretely, they showed that under certain technical assumptions, the dispersion relation of a NG boson is proportional to a positive integer power of momentum in the long-wavelength limit. Classifying NG bosons with an odd (even) power of momentum in the dispersion relation as type-I (type-II) and denoting their numbers as $n_{\text I}$ and $n_{\text{II}}$, respectively, they proved the inequality $n_{\mathrm{I}}+2n_{\mathrm{II}}\geq n_{\mathrm{BS}}$. In order to generalize their result to systems with only discrete translational symmetry, we basically just need to replace the spectral representation in their original paper by Eq.~\eqref{eq:S}. We shall nevertheless present some details of the argument.

Let us define the vectors $(\bm{A}_n)_b=\langle 0|\phi_b(0)|n\rangle$ and $\bm{v}_a=\bm{A}_n\langle n|\overline{j_a}|0\rangle$. Let in addition $p$ be the number of linearly independent vectors among the $\bm{v}_a$'s (obviously $p\leq n_{\mathrm{NG}}$), and set $\Delta n=n_{\mathrm{BS}}-p$. By construction, there is a set of coefficients $C_{a}^{\alpha}$ ($\alpha=1,\dotsc,\Delta n$) such that $C_{a}^{\alpha}\bm{v}_a=0$ and simultaneously $\mathrm{rank}\,C=\Delta n$. Using Eq.~\eqref{eq:S}, we obtain
\begin{align}
\notag
&iC_{a}^{\alpha}S_{ab}(\omega,\bm k)\simeq\\
&2\pi\sum_{n}\delta(\omega-\tfrac1\hbar\varepsilon_{n}(\bm{k}))\langle 0|C_{a}^{\alpha}\overline{j_a}_{(\bm{k})}|n,\bm{k}\rangle\langle n,\bm{k}|\phi_b(0)|0\rangle
\label{eq:analytic}
\end{align}
in the vicinity of $\bm{k}=\bm 0$. Here we dropped the contribution from massive modes and the second term of Eq.~\eqref{eq:S}, since they both vanish at $\bm{k}=\bm0$. Assuming along with Nielsen and Chadha the exponential decay of (expectation values of) commutators of local operators separated by a large spacelike interval, we infer that $S_{ab}(t,\bm{k})$ is an analytic function of $\bm{k}$. In other words, the support of $S_{ab}(\omega,\bm k)$ should be differentiable with respect to $\bm{k}$. As a consequence, for any $|n,\bm{k}\rangle$ that appears in Eq.~\eqref{eq:analytic} for some $\alpha$ and $b$, $\varepsilon_n(\bm{k})$ must be an analytic function of $\bm{k}$. We will call such NG modes type-C and denote  their number as $n_{\mathrm{C}}$.

As the next step, we observe that the $\Delta n$ vectors $C_{a}^{\alpha *}\bm{v}_a=\bm{A}_n\langle n|C_a^{\alpha*}\overline{j_a}|0\rangle$ must be linearly independent. Indeed, if this were not true, there would be a set of coefficients $\beta_{\alpha}$ such that $\beta_{\alpha}C_{a}^{\alpha*}\bm{v}_a$=0 and thus $\beta_{\alpha}C_{a}^{\alpha *}M_{ab}=0$, implying that $M$ has a zero eigenvalue, in contradiction with the assumption $\det M\neq0$. From the fact that the vectors $C_{a}^{\alpha *}\bm{v}_a$ are linearly independent, we can immediately conclude that the number of NG states $|n\rangle$ for which $\langle 0|C_a^{\alpha}\overline{j_a}|n\rangle\neq 0$ for some $\alpha$, must be at least $\Delta n$, hence $n_{\mathrm{C}}\geq \Delta n$. 

The remaining NG modes, for which $\langle 0|C_a^{\alpha}\overline{j_a}|n\rangle=0$ for all $\alpha$, will be called type-N and their number denoted accordingly as $n_{\text{N}}$. The above argument does not constrain in any way the dispersion relation of type-N NG bosons; in particular, it is allowed to be linear, $\varepsilon(\bm k)\propto|\bm k|$. Using the fact that, by construction, $n_{\text N}+n_{\text C}=n_{\text{NG}}$, we obtain the inequality
\begin{equation}
n_{\mathrm{N}}+2n_{\mathrm{C}}\geq n_{\mathrm{NG}}+\Delta n=n_{\mathrm{BS}}+(n_{\mathrm{NG}}-p)\geq n_{\mathrm{BS}}.
\label{eq:NC}
\end{equation}
Note that in Ref.~\cite{WatanabeBrauner:2011}, we actually proved the equality, $n_{\mathrm{N}}+2n_{\mathrm{C}}=n_{\mathrm{BS}}$, using a slightly different definition of the two types of the NG modes.

In fact, the derivation of the inequality~\eqref{eq:NC} does not make any use of rotational invariance which was implicitly, and somewhat unnecessarily, assumed in Ref.~\cite{Nielsen:1975hm}. Once this constraint is released, the classification of NG bosons can no longer be based on a mere power of the momentum $|\bm{k}|$ in the dispersion relation. A convenient generalization is to define as type-II those NG modes whose dispersion is analytic around $\bm{k}=\bm0$, and as type-I the remaining ones; this classification reduces to that of Nielsen and Chadha once rotational invariance is restored. We have thus shown in other words that every type-C NG boson is simultaneously type-II. Using the inequality \eqref{eq:NC}, we then find $n_{\mathrm{I}}+2n_{\mathrm{II}}\geq n_{\mathrm{N}}+2n_{\mathrm{C}}\geq n_{\mathrm{BS}}$, which reproduces the Nielsen--Chadha inequality.


\subsection{Implications for spontaneous breaking of translational invariance}
\label{subsec:discussion}

In the preceding subsections, we saw how the general theorems known in literature can be extended to account for the possibility that the ground state of the system has only a discrete translational symmetry. We are particularly interested in the situation when this arises as a result of a spontaneous breakdown of a continuous translational invariance. We will therefore devote this subsection to some discussion of systems with the symmetry breaking pattern
\begin{equation}
\mathcal{G}_{\text{int}}\times\mathcal{G}_{\rm C}^{(d)}\to\mathcal{G}_{\text{int}}'\times\mathcal{G}_{\rm D}^{(d')}\times\mathcal{G}_{\rm C}^{(d-d')},
\end{equation}
where $\mathcal{G}_{\text{int}}$ describes the internal symmetry group, which can also be (partially) spontaneously broken.

First, note that the Noether charge of translational invariance, that is, the momentum operator, commutes with all other broken charges $Q_a$. If we for a moment assume, as in the theorem of Sch\"afer~\emph{et al.}, that $\langle0|[Q_a,Q_b]|0\rangle=0$ for all pairs of broken internal symmetry generators, we immediately conclude that the number of NG bosons equals the number of broken generators. Therefore, there is one NG boson for each spontaneously broken translation generator as well as for each spontaneously broken generator of an internal symmetry.

In the general case, it is convenient to define the antisymmetric matrix $\rho$ by $i\rho_{ab}\equiv\displaystyle\lim_{\Omega\rightarrow \infty}\Omega^{-1}\langle0|[Q_a,Q_b]|0\rangle$ where $a,b=1,\dotsc,n_{\mathrm{BS}}$. In the recent paper~\cite{WatanabeBrauner:2011}, we proposed that the number of NG bosons is generally related to the rank of the matrix $\rho$ by
\begin{equation}
n_{\mathrm{BS}}-n_{\mathrm{NG}}=\tfrac12\mathrm{rank}\,\rho.
\label{eq:conjecture}
\end{equation}
The fact that translation generators commute with everything else now implies that $\mathrm{rank}\,\rho=\mathrm{rank}\,\rho_{\mathrm{int}}$, where $\rho_{\text{int}}$ is the analogous matrix defined using the generators of broken internal symmetries only. We can thus infer that the counting of NG bosons of spontaneously broken translational invariance is not affected by the presence of other spontaneously broken, internal symmetries: each spontaneously broken translation generator gives rise to exactly one NG boson. In Section~\ref{sec:supersolid}, we will discuss examples of both types of systems where $\mathrm{rank}\,\rho$ is zero and nonzero.

So far, we have tacitly assumed that the generators of space translations commute with one another as well as with other symmetry generators. Nevertheless, in presence of external fields, this property can be violated while still maintaining well-defined translational symmetry. Let us consider, for instance, a system of (interacting) charged particles in a uniform external magnetic field $\bm B$. Due to the explicit dependence of the vector potential on the coordinates, the action is no longer invariant under the usual translations. Instead, it is invariant under \emph{magnetic translations}, that is, translations combined with a gauge transformation of the electromagnetic field. The momentum operator $\bm P$ as a symmetry generator is then replaced by the operator $\bm P^B$, which satisfies
\begin{equation}
[P_i^{B},P_j^{B}]=-i\hbar q\epsilon_{ijk}B_kQ_0,\qquad
[P_i^{B},Q_0]=0,
\label{eq:crB}
\end{equation}
where $q$ is the charge of the particles and $Q_0$ the U(1) Noether charge, that is, their total number. Note that the rank of the matrix $\rho^B$ defined by $i\rho^B_{ij}\equiv\displaystyle\lim_{\Omega\rightarrow \infty}\Omega^{-1}\langle0|[P^B_i,P^B_j]|0\rangle$ equals two if the magnetic field as well as the expectation value of $Q_0$ is nonzero (otherwise the rank is zero). Eq.~\eqref{eq:conjecture} then suggests that in case magnetic translational invariance is spontaneously broken, there is one less NG boson than naively expected based on the number of broken translation generators. We will see an example of this in Section~\ref{sec:chargedphonon}~\footnote{As we observe in Section~\ref{sec:chargedphonon}, the magnetic translations are actually not uniform and one should therefore be very careful trying to apply our conjecture~\eqref{eq:conjecture} to this case. In fact, the magnetic translational symmetry is \emph{always} broken for systems with nonzero charge density.}.


\section{NG modes in a supersolid}
\label{sec:supersolid}

In this section, we will illustrate the general results obtained above on the example of NG modes in a (ferromagnetic) supersolid in $d$ spatial dimensions. We will employ the mean-field approximation at zero temperature. 


\subsection{Formalism}
\subsubsection{Definition of the model and its symmetry}

Let us consider the class of theories, defined by the following Lagrangian,
\begin{equation}
\begin{split}
L[\Psi]=\int d^d\bm{x}\,\Psi^{\dagger}_i(x)\left[i\hbar\partial_t+\frac{\hbar^2\bm\nabla^2}{2m}+\mu\right]\Psi_i(x)\\
-\frac{1}{2}\int d^d\bm{x}\,d^d\bm{y}\,\Psi^{\dagger}_i(x)\Psi_i(x)V(\bm{x}-\bm{y})\Psi^{\dagger}_j(y)\Psi_j(y),
\end{split}
\label{eq:L}
\end{equation}
where $\Psi=(\Psi_1,\dotsc,\Psi_N)$ is an $N$-component complex scalar field, and the potential $V(\bm{x})=V(-\bm{x})$ represents a finite-range interaction. The symmetry of the Lagrangian is obviously U$(N)\times\mathcal{G}_{\rm C}^{(d)}$. The simplest case $N=1$ corresponds to the usual Bose--Einstein condensate (BEC) without internal degrees of freedom, while the widely discussed $N=2$ case can be understood as a spin-1/2 BEC (see, for instance, Ref.~\cite{Kasamatsu:2005}).

It is convenient to analyze the model using dimensionless variables; this makes the results universally applicable to different physical systems, as long as they share the same set of symmetries and degrees of freedom. We first introduce a characteristic length scale of the potential, $a$, in terms of which the strength of the interaction can be measured by the parameter $V_0$, defined by $a^dV_0=\int d^d\bm{x}\,V(\bm{x})$. Next, we rescale the spacetime coordinates as $t\to t\hbar/\mu$ and $\bm x\to\bm xa$. (For conventional reasons we use the same symbols for the new, dimensionless coordinates; as we do so systematically from now on, no confusion can arise.) Finally, we introduce the dimensionless potential, $v(\bm{x})=V(\bm{x})/V_0$, and the field $\psi(x)=\Psi(x)\sqrt{V_0 a^d/\mu}$. The dimensionless Lagrangian, $L[\psi]=L[\Psi] V_0/\mu^2$, thus becomes
\begin{equation}
\begin{split}
L[\psi]=\int d^d\bm{x}\,\psi^{\dagger}_i(x)\left(i\partial_t+\frac{\bm\nabla^2}{2g}+1\right)\psi_i(x)\\
-\frac{1}{2}\int d^d\bm{x}\,d^d\bm{y}\,\psi^{\dagger}_i(x)\psi_i(x)v(\bm{x}-\bm{y})\psi^{\dagger}_j(y)\psi_j(y),
\end{split}
\label{eq:L2}
\end{equation}
where $g$ is a dimensionless coupling parameter defined as $g=\mu(\hbar^2/ma^2)^{-1}$.


\subsubsection{Classical field}
\label{subsec:classical}

The first step in the mean-field analysis is to find the ground state, that is, the \emph{static} classical field configuration $\psi_{0}(\bm{x})$ which minimizes the energy,
\begin{equation}
\begin{split}
E[\psi]=\int d^d\bm{x}\left[\frac{1}{2g}\bm{\nabla}\psi^*_i(\bm{x})\cdot\bm{\nabla}\psi_i(\bm{x})-\psi^*_i(\bm{x})\psi_i(\bm{x})\right]\\
+\frac{1}{2}\int d^d\bm{x}\,d^d\bm{y}\,\psi^*_i(\bm{x})\psi_i(\bm{x})v(\bm{x}-\bm{y})\psi^*_j(\bm{y})\psi_j(\bm{y}).
\end{split}
\label{eq:E}
\end{equation}
To that end, we exploit the symmetry of the problem and cast the spinor $\psi_0(\bm x)$ in a ``canonical'' form. One should nevertheless be careful since the field in general depends on the coordinates. Note that it can always be written as $\psi(\bm x)=U(\bm x)\psi'(\bm x)$, where $U(\bm x)$ is a unitary matrix and $\psi'_{i}(\bm{x})=\delta_{i1}\bar{\psi}(\bm{x})$ with real $\bar\psi(\bm x)$. As long as the field $\psi(\bm x)$ does not contain any vortices, both $U(\bm x)$ and $\bar\psi(\bm x)$ are differentiable. It then follows that
\begin{equation}
E[\psi]-E[\psi']=\sum_{i=1}^N\int d^d\bm{x}\,\frac{\bar{\psi}(\bm{x})^2|\bm{\nabla}U_{i1}(\bm{\bm{x}})|^2}{2g}\geq 0.
\end{equation}
The equality occurs only if $U_{i1}$ is coordinate-independent. Thus, by a global U$(N)$ rotation, $\psi_{0}(\bm{x})$ can always be chosen as $\psi_{0i}(\bm{x})=\delta_{i1}\bar{\psi}_{0}(\bm{x})$ with a real field $\bar{\psi}_0(\bm{x})$. Field configurations with vortices tend to have a higher energy and we therefore assume we can discard this possibility.

One (nontrivial) stationary point of the energy functional~\eqref{eq:E} is always given by a uniform field. Due to our normalization, $\int d^d\bm{x}\,v(\bm{x})=1$, this corresponds to $\bar\psi_0(\bm{x})=1$ with the energy density $-1/2$. The uniform solution indeed minimizes energy for small $g$. However, when $g$ becomes large, the energy cost of creating spatial modulation of $\psi(\bm x)$ is suppressed. Depending on the detailed structure of the potential $v(\bm x)$, this cost can then be overwhelmed by the energy gain from the nonlocal interaction. As a result, inhomogeneous field configurations may become energetically favorable.


\subsubsection{Spontaneous symmetry breaking}
\label{subsec:SSB}

As soon as the classical field $\psi_0(\bm x)$ assumes a nonzero value, the symmetry of the Lagrangian is spontaneously broken. Owing to the fact that the spinor $\psi_{0}(\bm x)$ points in the same direction everywhere in space, $\psi_{0i}(\bm{x})=\delta_{i1}\bar{\psi}_{0}(\bm{x})$, the internal U$(N)$ symmetry is broken down to its U$(N-1)$ subgroup; $2N-1$ of its generators are thus broken. Therefore, when $\psi_0(\bm x)$ is homogeneous the residual symmetry is U$(N-1)\times\mathcal G_{\rm C}^{(d)}$ and $n_{\text{BS}}=2N-1$. We will call this solution the \emph{superfluid phase}. On the other hand, if $\psi_0(\bm x)$ is inhomogeneous and periodically modulated in all spatial directions (the \emph{supersolid phase}), the symmetry of the ground state is merely U$(N-1)\times\mathcal{G}_{\rm D}^{(d)}$ and $n_{\mathrm{BS}}=2N-1+d$.

To predict the number of NG modes, we have to evaluate the rank of the matrix $\rho$ in Eq.~\eqref{eq:conjecture}. To that end, note that the $2N-1$ broken internal symmetry generators fall into three classes: the generator of phase transformations of the first component of the spinor, $(T_1)_{jk}=\delta_{j1}\delta_{k1}/2$; the $N-1$ real generators $(T_{\ell}^{\rm R})_{jk}=(\delta_{k1}\delta_{j\ell}+\delta_{j1}\delta_{k\ell})/2$; the $N-1$ pure imaginary generators $(T_{\ell}^{\rm I})_{jk}=i(\delta_{k1}\delta_{j\ell}-\delta_{j1}\delta_{k\ell})/2$, where $\ell=2,3,\dotsc,N$. Using these matrices, the broken generators can be represented on the Hilbert space of the system by the operators $Q_a=\int d^d\bm{x}\,\psi_i^{\dagger}(x)(T_a)_{ij}\psi_j(x)$.  Within the mean-field approximation, the fluctuations are neglected and one easily derives
\begin{equation}
\begin{split}
\frac{1}{\Omega}\langle 0|[Q_{\ell}^{\rm R},Q_{\ell'}^{\rm I}]|0\rangle&=\frac{i\delta_{\ell\ell'}}{\Omega}\langle 0|Q_1|0\rangle\\
&=\dfrac{i\delta_{\ell\ell'}}{2\Omega_{\mathrm{uc}}}\int_{\mathrm{uc}}d^d\bm{x}\,\bar{\psi}_0(\bm{x})^2
\end{split}
\end{equation}
for $\ell,\ell'=2,\dotsc,N$ and $\langle 0|[Q_a,Q_b]|0\rangle=0$ for other combinations. We thus conclude that $(1/2)\mathrm{rank}\,\rho=N-1$.

In the case $N=1$ (usual BEC), we can now use the theorem of Sch\"afer~\emph{et al.} to assert that $n_{\mathrm{NG}}=n_{\mathrm{BS}}=1$ in the superfluid phase and $n_{\mathrm{NG}}=n_{\mathrm{BS}}=1+d$ in the supersolid phase. For $N\geq 2$, Eq.~\eqref{eq:conjecture} implies that $n_{\mathrm{NG}}=N$ in the superfluid phase and $n_{\mathrm{NG}}=N+d$ in the supersolid phase. To be consistent with the Nielsen--Chadha theorem, at least $N-1$ of the NG modes must be type-II. This is in agreement with the result of Ref.~\cite{Andersen:2005yk} which analyzed the relativistic version of the model with a local interaction, $V(\bm x)\propto\delta^d(\bm x)$. The detailed nature of the NG modes in our model will be investigated in the following subsection.


\subsubsection{Elementary excitations}

In the mean-field approximation, the excitation spectrum is found simply by expanding the Lagrangian to second order in fluctuations around the classical field. We parameterize the field as $\psi(x)=\psi_{0}(\bm{x})+\phi(x)+i\varphi(x)$ and plug this back into Eq.~\eqref{eq:L2}. Up to second order in the fluctuation fields $\phi,\varphi$, the Lagrangian reads $L[\psi]=L[\psi_0]+L_2[\phi,\varphi;\psi_0]+\cdots$, where
\begin{equation}
\begin{split}
L_2[\phi,\varphi;\psi_0]=&\int d^d\bm{x}\,\Bigl\{\phi_i(x)\mathcal{K}_x\phi_i(x)+\varphi_i(x)\mathcal{K}_x\varphi_i(x)\\
&+\bigl[\varphi_i(x)\partial_t\phi_i(x)-\phi_i(x)\partial_t\varphi_i(x)\bigr]\Bigr\}\\
&-2\int d^d\bm{x}\,d^d\bm{y}\,\phi_1(x)w(\bm{x},\bm{y})\phi_1(y)
\end{split}
\label{eq:Lbilin}
\end{equation}
up to a term which is a total time derivative. In the above equation, $\mathcal{K}_x=(\bm\nabla^2/2g)+1-u(\bm{x})$ is the kinetic term, $u(\bm{x})=\int d^d\bm{y}\,v(\bm{x}-\bm{y})|\bar{\psi}_0(\bm{y})|^2$ is the effective periodic mean-field potential due to the background $\psi_0(\bm{x})$, and $w(\bm{x},\bm{y})=\bar{\psi}_0(\bm{x})v(\bm{x}-\bm{y})\bar{\psi}_0(\bm{y})$ captures the effects of the non-local interaction.

Physically, the fluctuations $\{\phi_i\}_{i=2}^{N}$ and $\{\varphi_i\}_{i=1}^{N}$ correspond to NG fields of the spontaneously broken internal symmetry. Namely, the $\{\phi_i\}_{i=2}^{N}$ are obtained from the classical field $\psi_0(\bm x)$ by real rotations, generated by the matrices $T^{\rm I}_i$, while the $\{\varphi_i\}_{i=1}^{N}$ are induced by symmetric unitary transformations, generated by the matrices $T^{\rm R}_i$ (or $T_1$ in case of $i=1$). On the other hand, the $\phi_1(x)$ mode corresponds to displacements of the classical field configuration, $\psi_{0}(\bm{x}+\bm{\xi}(x))-\psi_{0}(\bm{x})\simeq\bm{\xi}(x)\cdot\bm{\nabla}\psi_0(\bm{x})$, as well as to density fluctuations. This can be seen by noting the role of $\phi_1(x)$ in the density--density correlation function,
\begin{equation}
\begin{split}
&\langle0|T\{\delta n(x)\delta n(y)\}|0\rangle\\
&\simeq4\bar{\psi}_0(\bm{x})\bar{\psi}_0(\bm{y})\langle0|T\{\phi_1(x)\phi_1(y)\}|0\rangle+\cdots,
\end{split}
\end{equation}
where $\delta n(x)=\psi_i^{\dagger}(x)\psi_i(x)-\langle0|\psi_i^{\dagger}(x)\psi_i(x)|0\rangle$, and the ellipsis stands for terms of higher order in the fluctuation fields.

The dispersion relations of the various excitation branches are determined formally by a diagonalization of the Lagrangian~\eqref{eq:Lbilin}. In practice, we accomplish this task by solving the Euler--Lagrange equations for the fluctuation fields,
\begin{equation}
\begin{split}
\partial_t\varphi_i(x)&=\mathcal{K}_x\phi_i(x)-2\delta_{i1}\int d^d\bm{y}\,w(\bm{x},\bm{y})\phi_1(y),\\
\partial_t\phi_i(x)&=-\mathcal{K}_x\varphi_i(x).
\end{split}
\label{eq:Sch}
\end{equation}
Note that the equations of motion for $\phi_i$ and $\varphi_i$ are coupled; we will see the importance of this coupling later. Based on the residual discrete symmetry, $\mathcal{G}_{\rm D}^{(d)}$, we can assume the normal modes of $\phi_i,\varphi_i$ to have the Bloch form, that is,
\begin{equation}
\phi_{i,n}(\bm{k},x)=\sum_{\bm{G}}\phi_{i,n,\bm{G}}(\bm{k})\,e^{i(\bm{k}+\bm{G})\cdot\bm{x}-i\omega_n(\bm{k})t},
\end{equation}
(analogously for $\varphi_i$) with a band index $n$ and a crystal momentum $\bm{k}\in\text{FBZ}$. The solution $\omega_n(\bm{k})$ represents the allowed excitation spectrum. In particular, those modes which satisfy $\omega_n(0)=0$ are NG modes, assuming they couple to the spontaneously broken currents.


\subsection{Numerical results}

As a concrete illustration of the general setup, we choose the simplest model for a supersolid, defined by the the step-function potential $V(\bm{x})=(V_0/\pi)\theta(a-|\bm{x}|)$ in two spatial dimensions~\cite{Pomeau:1994}. Below, we summarize the results of our numerical computations. In order not to obscure the physics with technicalities, we describe some details of the methods we used in Appendix~\ref{app:methods}. In particular, in Appendix~\ref{app:minimize} we explain how the classical field $\bar\psi_0(\bm x)$ was found, and in Appendix~\ref{app:band} how the band structures presented in Figs.~\ref{fig2a} and \ref{fig2b} were obtained.


\subsubsection{Superfluid phase}

\begin{figure}
\begin{minipage}{0.6\hsize}
\subfigure[]{\includegraphics[width=\textwidth]{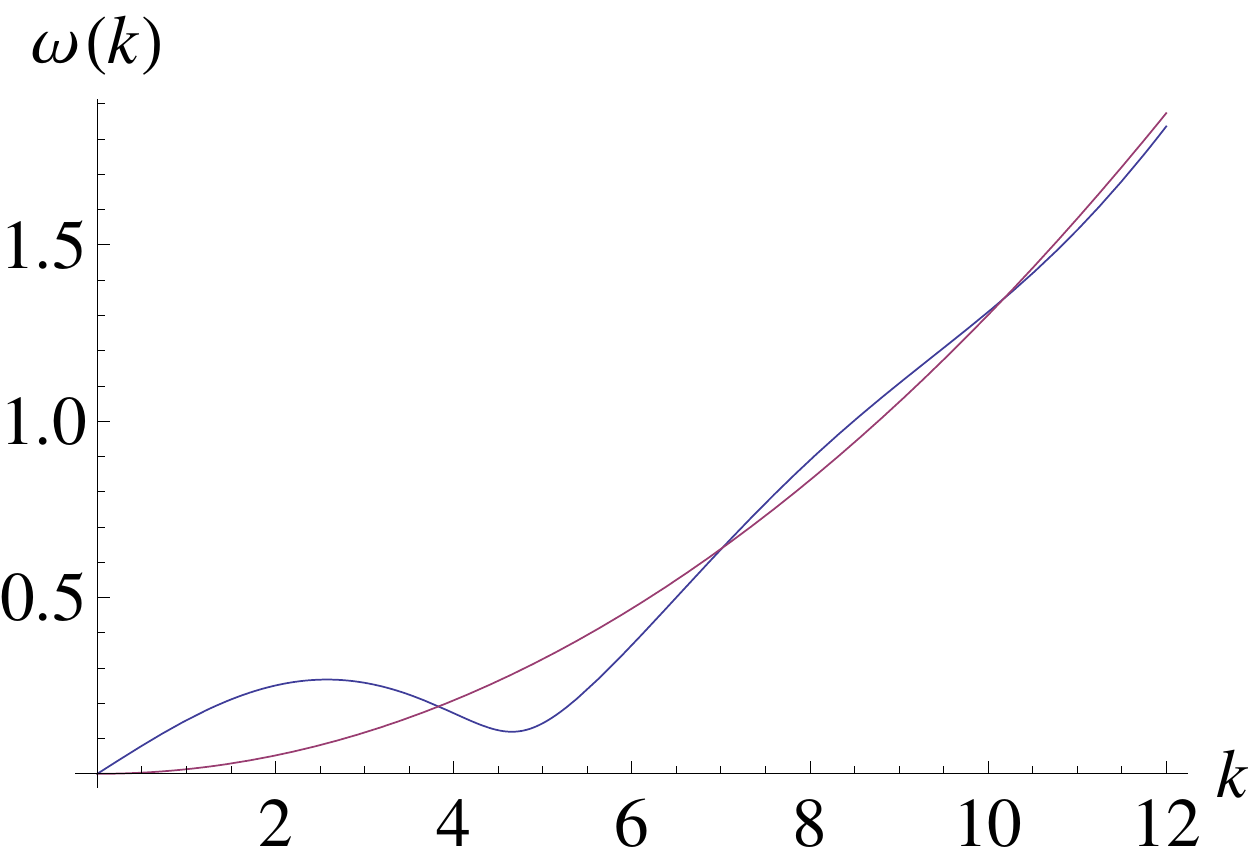}\label{fig1a}}
\end{minipage}
\begin{minipage}{0.36\hsize}
\subfigure[]{\includegraphics[width=\textwidth]{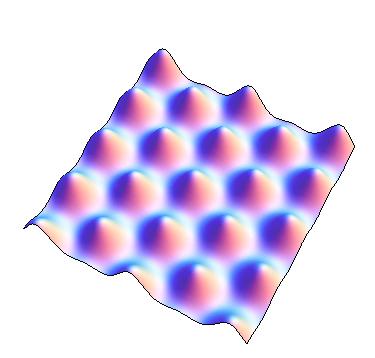}\label{fig1b}}
\end{minipage}
\caption{(color online). (a) The dispersion relations of NG bosons in the superfluid phase at $g=g_0-0$. The linear branch is the usual Bogoliubov mode, while the quadratic branch (with $N-1$ fold degeneracy) can be interpreted as a ferromagnetic magnon. The dispersion of the Bogoliubov mode has a roton-like minimum around $|\bm k|\approx4.80$, which corresponds to the magnitude of the reciprocal lattice vector at $g=g_0+0$. (b) Three-dimensional plot of the classical field $\bar{\psi}_0(\bm{x})$ in the supersolid phase. The energy functional is minimized by a hexagonal (triangular) lattice with the lattice constant approximately $1.50$ at $g=g_0+0$. Note that $\bar{\psi}_0(\bm{x})$ does not have any nodes.}
\end{figure}

For $g<g_0\approx38.4$, the energy~\eqref{eq:E} is minimized by the homogeneous solution $\bar\psi_0(\bm x)=1$~\footnote{In Ref.~\cite{Sepulveda:2010}, a model equivalent to ours with $N=1$ was analyzed and the critical coupling was found to be approximately $41.6$. The reason for the discrepancy is twofold. First, the authors of Ref.~\cite{Sepulveda:2010} studied the model on a square domain with periodic boundary conditions, which somewhat disfavors the supersolid state with a hexagonal structure and thus shifts the transition towards stronger couplings. Second, in Ref.~\cite{Sepulveda:2010} the energy functional was minimized with a fixed particle number, whereas we fixed the chemical potential instead. As a consequence, the average particle density has a tiny discontinuity at the transition point: while in the superfluid phase it is exactly one by normalization, in the supersolid phase it is approximately $1.06$.}. In this superfluid phase, the Schr\"odinger equations~\eqref{eq:Sch} are solved straightforwardly by going to the energy--momentum space, taking the form $A_i(\bm{k})\begin{pmatrix}\phi_i(\bm{k}), \varphi_i(\bm{k})\end{pmatrix}^T=0$ (separately for $i=1,\dotsc,N$), where
\begin{equation}
A_i(\bm{k})
=\begin{pmatrix}
(\bm k^2/2g)+2\delta_{i1}v(\bm{k}) & -i\omega(\bm{k})\\
i\omega(\bm{k}) & \bm k^2/2g
\end{pmatrix}.
\end{equation}
Here $\phi_i(\bm{k})=\int d^2\bm{x}\,\phi_i(0,\bm{x})e^{-i\bm{k}\cdot\bm{x}+i\omega(\bm{k})t}$ [analogously for $\varphi_i(\bm k)$] and $v(\bm{k})=\int d^2\bm{x}\,v(\bm{x})e^{-i\bm{k}\cdot\bm{x}}=2J_1(|\bm k|)/|\bm k|$ [$J_n(x)$ is the Bessel function]. From the condition $\det A_i(\bm{k})=0$ we find the quasiparticle dispersion relation, $\omega_i(\bm k)=\sqrt{(\bm k^2/2g)\left[(\bm k^2/2g)+2\delta_{i1}v(\bm k)\right]}$. Around $\bm k=\bm0$, the dispersions and eigenvectors corresponding to the individual modes are expanded as 
\begin{align}
\omega_1(\bm k)&\simeq|\bm k|/\sqrt{g},
&&\begin{pmatrix}
\phi_1,\varphi_1
\end{pmatrix}
\simeq
\begin{pmatrix}
0,1
\end{pmatrix},\\
\omega_j(\bm k)&\simeq\bm k^2/2g,
&&\begin{pmatrix}
\phi_j,\varphi_j
\end{pmatrix}
\simeq\tfrac1{\sqrt2}
\begin{pmatrix}
1,-i
\end{pmatrix},\qquad(j\geq 2),
\notag
\end{align}
to leading order in momentum. We illustrate these dispersion relations in Fig.~\ref{fig1a} for the coupling set to the transition point, $g=g_0-0$. In summary, there is one type-I Bogoliubov mode and $N-1$ type-II modes with a quadratic dispersion and a circular polarization, which can be interpreted as ferromagnetic magnons (precession modes)~\cite{Ho:1998,*Ohmi:1998}. This conclusion is in exact agreement with the prediction we made in Section~\ref{subsec:SSB}.

It is worth emphasizing that while at $\bm k=\bm0$ the Bogoliubov mode corresponds to a pure phase fluctuation, $\varphi_1$, at nonzero momentum it is given by a mixture of $\varphi_1$ and $\phi_1$. In other words, the Bogoliubov mode generates not only a phase modulation but also a density modulation of the condensate $\psi_0$. This is why we can associate it with a density wave and detect it in Bragg spectroscopy experiments~\cite{Steinhauer:2002}. However, this very fact has also caused some confusion regarding the independence of the Bogoliubov mode and the longitudinal phonon originating from the crystalline order in the supersolid phase. We will shortly see that these two are, indeed, distinct modes.

Finally, let us remark that the above discussed mixing can be traced back to the term $\psi^{\dagger}\partial_t\psi$ in the Lagrangian~\eqref{eq:L2}. A similar term also arises in the analogous relativistic model where it is induced by the chemical potential~\cite{Andersen:2005yk}. On the other hand, in a Lorentz-invariant model where SSB is triggered by a ``wrong sign'' of the mass term of the field, no such mixing would arise, and the $\phi_i$ and $\varphi_i$ modes would be completely decoupled.


\subsubsection{Supersolid phase}

For $g>g_0$, periodic spatial modulation of the order parameter $\bar\psi_0(\bm x)$ is energetically favored. The transition at $g=g_0$ is of first order within the mean-field approximation~\cite{Pomeau:1994}, and can be roughly understood as Bose--Einstein condensation of field modes with momentum determined by the reciprocal lattice vectors. This interpretation is supported by the roton-like minimum in the dispersion relation of the Bogoliubov mode shown in Fig.~\ref{fig1a}. Numerical computation, following the strategy outlined in Appendix~\ref{app:minimize}, shows that the solution with the lowest energy has the form of a hexagonal lattice, plotted in Fig.~\ref{fig1b}.

\begin{figure}
\subfigure[]{\includegraphics[width=0.4\textwidth]{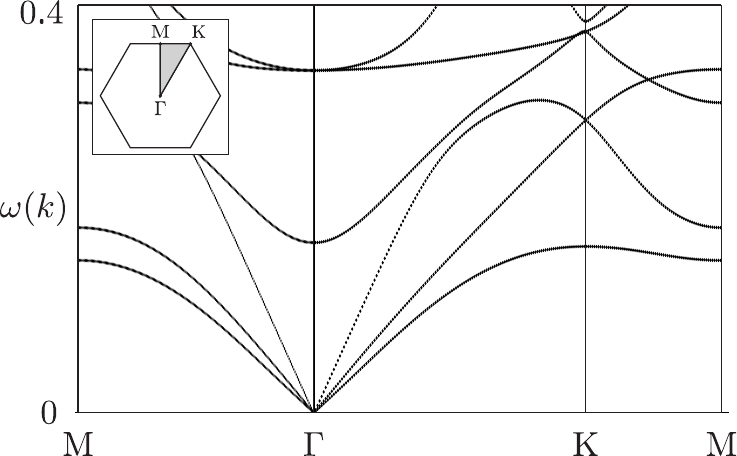}\label{fig2a}}
\subfigure[]{\includegraphics[width=0.4\textwidth]{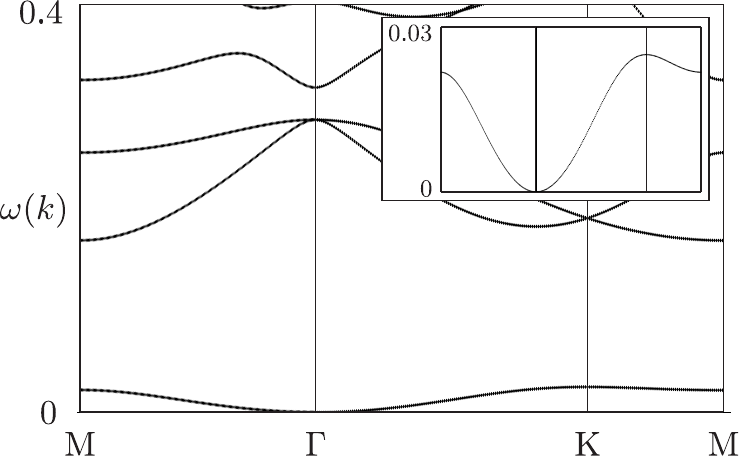}\label{fig2b}}
\caption{The numerical result for dispersion relations in the supersolid phase at $g=g_0+0$. Panel (a) displays the excitation branches in the $\phi_1$, $\varphi_1$ sector. We can see three linear dispersions around the $\Gamma$ point, which correspond to the Bogoliubov mode and two phonons. The inset shows the irreducible part of FBZ. Panel (b) displays the excitation branches in the $\phi_i$, $\varphi_i$ ($i\geq 2$) sector. The dispersion relation of the gapless mode is quadratic at low momentum, as can be seen in the inset which zooms in the vertical axis. }
\end{figure}

By solving the coupled Schr\"odinger equations~\eqref{eq:Sch} using the method explained in Appendix~\ref{app:band}, we found the band structures in the supersolid  phase at $g=g_0+0$, shown in Figs.~\ref{fig2a} and~\ref{fig2b}. This result was obtained using the first 73 $\bm G$ vectors around the origin of the reciprocal lattice. The convergence of the result was checked by a comparison with one using only 61 $\bm G$ vectors.

In the $i=1$ sector, three of the excitation branches are linear around the $\Gamma$ point (the origin of the $\bm{k}$ space), corresponding to a longitudinal phonon, a transverse phonon, and a Bogoliubov mode. On the other hand, each $i\geq 2$ sector contains just one massless mode with a quadratic dispersion relation---the magnon. In summary, there are in total $N+d$ NG modes, $N-1$ of which are type-II, exactly as predicted in Section~\ref{subsec:SSB}. 

In addition, we have checked numerically that the dispersion relations of all the gapless modes are isotropic at the $\Gamma$ point; the NG modes of the model will thus be described by a rotationally invariant low-energy effective field theory. Our findings are consistent with previous works for the case $N=1$, based on the effective Lagrangian method~\cite{Son:2005,*Josserand:2007,*Ye:2008}. A low energy effective field theory describing simultaneously the Bogoliubov mode and the phonons has recently also been worked out, in a different context, in Ref.~\cite{Cirigliano:2011tj}.

We would like to emphasize that while the specific results presented in Figs.~\ref{fig2a} and~\ref{fig2b} rely on the mean-field approximation, the qualitative conclusion that the Bogoliubov mode and the longitudinal phonon are distinct physical excitations is exact. Following the general argument of Section~\ref{subsec:discussion}, in the case of $N=1$ it is based on the theorem of Sch\"afer~\emph{et al.} and thus can be considered rigorously proven. For $N\geq2$, it is strongly supported by the conjectured equality~\eqref{eq:conjecture}.


\section{Phonons in a charged crystal in external magnetic field}
\label{sec:chargedphonon}

As anticipated in the discussion in Section~\ref{subsec:discussion}, the spectrum of NG modes can become more complicated once the generators of space translations do not commute with each other as well as with generators of other, internal symmetries. To illustrate this subtlety, we shall in this section discuss briefly a simple quantum-mechanical model for a charged crystal.


\subsection{Quantum-mechanical model}

The Hamiltonian for $N$ particles with the common charge $q$ in non-relativistic quantum mechanics is given by
\begin{equation}
H=\sum_{i=1}^N\frac{[\bm{p}_i-q\bm{A}(\bm{x}_i)]^2}{2m}+\dfrac{1}{2}\sum_{i\neq j}v(\bm{x}_i-\bm{x}_j),
\end{equation}
where $\bm{A}(\bm{x}_i)$ is a vector potential which determines the external magnetic field $\bm B(\bm x)$ via $\bm{\nabla}\times\bm{A}(\bm{x})=\bm{B}(\bm{x})$, and $v(\bm{x})$ is a repulsive interaction potential which is assumed to be invariant under space inversion, $v(\bm{x})=v(-\bm{x})$. This model, with the Coulomb potential $v(\bm{x})=e^2/|\bm{x}|$, has been used in the context of the Wigner solid~\cite{Fukuyama:1975}. Nevertheless, since our goal here is to discuss the consequences of SSB in the presence of a magnetic field, we will assume that the potential $v(\bm x)$ has a finite range.

The canonical coordinate and momentum $x_{i\alpha},p_{j\beta}$ are required to satisfy the usual commutation relation. (Within this section, we use Greek letters to denote spatial indices.) Because of the vector potential, the operator of total momentum $\sum_{i=1}^N\bm{p}_i$ no longer commutes with the Hamiltonian. When the magnetic field is uniform, we can instead define the operator $\bm{\rho}_i\equiv\bm{p}_i-q\left[\bm{A}(\bm{x_i})-\bm{B}\times\bm{x}_i\right]$. The sum, $P^B_\alpha=\sum_{i=1}^N\rho_{i\alpha}$, is conserved and plays the role of the generator of spatial translations ($[iP_{\alpha}^B,x_{i\beta}]=\hbar\delta_{\alpha\beta}$). At the same time, it satisfies the commutation relation~\eqref{eq:crB} with $Q_0=N$.


\subsection{Phonon spectrum}

Suppose that we have found a stable crystal configuration where the $i$-th particle is localized around the lattice node, $\bm{x}_i=\bm{R}_i$. The spectrum of oscillations of the crystal lattice can be determined by resorting to the harmonic approximation, in which the Hamiltonian takes the form $H=\sum_{i=1}^N[\bm{p}_i-q\bm{A}(\bm{x}_i)]^2/(2m)+(m/2)\sum_{i,j=1}^NW_{ij}^{\alpha\beta}u_{i\alpha}u_ { j\beta}$, where $\bm{u}_i\equiv \bm{x}_i-\bm{R}_i$ and
\begin{equation}
\begin{split}
mW_{ij}^{\alpha\beta}=
\begin{cases}-\partial_{\alpha}\partial_{\beta}v\Big|_{\bm{x}=\bm{R}_{ij}}, &(i\neq j),\\
\displaystyle\sum_{\substack{\ell=1\\ \ell\neq i}}^N\partial_{\alpha}\partial_{\beta}v\Big|_{\bm{x}=\bm{R}_{i\ell}}, &(i=j).
\end{cases}
\end{split}
\end{equation}
(We have also introduced the notation $\bm R_{ij}=\bm R_i-\bm R_j$.) Let us now assume that the total particle number $N$ extends to infinity or that periodic boundary conditions are imposed. Then the Heisenberg equations of motion---unlike the Hamiltonian itself---have the lattice translational symmetry, since only the magnetic field $\bm{B}$ appears in them. Thus, by using the Fourier components, $W_{\bm{k}}^{\alpha\beta}=\sum_iW^{\alpha\beta}_{ij}e^{-i\bm{k}\cdot\bm{R}_{ij}}$ for any fixed $j$, and $u_{\bm{k}\alpha}(t)=\sum_iu_{i\alpha}e^{i\omega(\bm{k})t-i\bm{k}\cdot\bm{R}_i}$, we simplify the equations to the form $M_{\alpha\beta}(\bm{k})u_{\bm{k}\beta}=0$ with $M_{\alpha\beta}(\bm{k})=\delta_{\alpha\beta}\omega(\bm{k})^2-i\epsilon_{\alpha\beta\gamma}\omega(\bm{k})b^{\gamma}-W_{\bm{k}}^{\alpha\beta}$ and $b^{\gamma}=qB^{\gamma}/m$. As before, the condition $\det M(\bm{k})=0$ determines the dispersion relations of the normal modes of the crystal lattice.

For the sake of brevity, we shall consider the simplest, albeit unrealistic, model---the \emph{isotropic} crystal in three spatial dimensions, where
\begin{equation}
W^{\alpha\beta}_{\bm k}=\left(\delta^{\alpha\beta}-\frac{k^\alpha k^\beta}{\bm k^2}\right)\omega_t(|\bm k|)^2+\frac{k^\alpha k^\beta}{\bm k^2}\omega_\ell(|\bm k|)^2.
\label{isotropic}
\end{equation}
In the absence of the magnetic field, the functions $\omega_t(|\bm k|)$ and $\omega_\ell(|\bm k|)$ give the dispersion relations of transverse and longitudinal phonons, respectively, and it is therefore natural to assume both of them to be linear at low momentum. The determinant of $M(\bm k)$ is easily found,
\begin{align}
\notag
\det M=&(\omega^2-\omega_t^2)^2(\omega^2-\omega_\ell^2)-\omega^2(\omega^2-\omega_t^2)\bm b^2\sin^2\theta\\
&-\omega^2(\omega^2-\omega_\ell^2)\bm b^2\cos^2\theta,
\end{align}
where $\theta$ is the angle between the vectors $\bm k$ and $\bm b$, and we have for simplicity omitted the argument $\bm k$ where necessary. At low momentum, the dispersion relations of the three normal modes can be evaluated analytically,
\begin{align}
\notag
\omega&\simeq|\bm b|, &&\text{gapped mode},\\
\label{cyclotron}
\omega&\simeq\sqrt{\omega_t^2\sin^2\theta+\omega_\ell^2\cos^2\theta}, &&\text{type-I NG mode},\\
\notag
\omega&\simeq\frac{\omega_t^2\omega_\ell}{|\bm b|\sqrt{\omega_t^2\sin^2\theta+\omega_\ell^2\cos^2\theta}}, &&\text{type-II NG mode}.
\end{align}
In order to determine the polarization of the modes, that is, the orientation of the vector $\bm u_{\bm k}$, it is convenient to rewrite the equation of motion in a vector form
\begin{equation}
\omega^2\bm u-i\omega\bm u\times\bm b-(\omega_t^2P_t\bm u+\omega_\ell^2P_\ell\bm u)=\bm0,
\label{eom}
\end{equation}
where $P_{t,\ell}$ are projectors to transverse/longitudinal directions with respect to momentum $\bm k$, defined by the respective terms in Eq.~\eqref{isotropic}. For the gapped mode, the terms in Eq.~\eqref{eom} proportional to $\omega_{t,\ell}^2$ can be neglected, which leads to the condition $\bm u=i\bm u\times\bm b/|\bm b|$. In words, this mode is circularly polarized in the plane perpendicular to the magnetic field, in the opposite direction than the usual cyclotron motion. For the type-I NG mode, we likewise neglect terms in Eq.~\eqref{eom} proportional to $\omega^2$ as well as $\omega_{t,\ell}^2$ to arrive at the constraint $\bm u\times\bm b=\bm0$, which simply means that this mode is always polarized along the direction of the magnetic field. Finally, for the type-II NG mode we neglect the term in Eq.~\eqref{eom} proportional to $\omega^2$ and obtain the condition $\bm u\times\bm b=(i/\omega)(\omega_t^2P_t\bm u+\omega_\ell^2P_\ell\bm u)$. 

Let us for illustration consider some special cases. When the momentum is parallel to the magnetic field ($\theta=0$), the energies of the modes given in Eq.~\eqref{cyclotron} reduce to $|\bm b|$, $\omega_\ell$, and $\omega_t^2/|\bm b|$, respectively. The type-II NG excitation is circularly polarized in the plane perpendicular to $\bm b$, in the opposite direction than the gapped mode. This is in a nice analogy to the type-II NG boson in ferromagnets---the magnon---where the circular polarization corresponds to the Larmor precession of the electron spin. Note that in the limit when $\omega_t=\omega_\ell$, the same conclusions hold regardless of the orientation of the momentum.

Second, when the momentum is perpendicular to the magnetic field ($\theta=\pi/2$), the dispersion relations in Eq.~\eqref{cyclotron} become $|\bm b|$, $\omega_t$, and $\omega_t\omega_\ell/|\bm b|$, respectively. The type-II NG mode is now polarized elliptically in the plane transverse to the magnetic field. The ratio of the principal axes of the ellipse, oriented along and perpendicularly to the momentum, is given by $\omega_t/\omega_\ell$. 

In the general case, as already mentioned above, the polarization of the gapped mode is circular and transverse to the magnetic field, while that of the type-I NG mode is linear and aligned with the magnetic field, irrespective to the orientation of the momentum. Only the polarization of the type-II NG excitation is affected by the momentum.

In any case, there are only two (that is, $d-1$) gapless phonons, and one of them has a quadratic dispersion relation. This can be explained by the combination of the conjecture~\eqref{eq:conjecture} and the Nielsen--Chadha theorem. The non-commutativity of the components of the generator of magnetic translations suggests that there is one less phonon than one would naively expect. Then, to satisfy the Nielsen--Chadha theorem, at least one of the phonons must be type-II.


\subsection{Possible application}

The second-quantized version of the above model can be used to describe the rapidly rotating BEC of cold atoms. When viewed from the co-rotating frame, the Hamiltonian of the system picks an additional term, equal to $-\bm{L}\cdot\bm{\Omega}$ (coupling of the angular momentum $\bm L$ to the angular velocity $\bm{\Omega}$). It then has formally the same form as the Hamiltonian of charged particles in the external magnetic field, provided the symmetric gauge is used and the replacement $\bm\Omega\to q\bm B/(2m)$ made~\cite{Pethick:2008}. Therefore, the Hamiltonian of the rotating Bose gas has a two-dimensional continuous magnetic translational symmetry, in addition to the U$(1)$ global symmetry corresponding to the conservation of particle number, as long as the trapping potential (modified by $\bm{\Omega}$) can be neglected.

The formation of a triangular vortex lattice has been experimentally observed~\cite{Abo:2001}, which implies spontaneous breaking of the continuous magnetic translational invariance. The collective mode due to the lattice distortion, called the Tkachenko mode, has also been experimentally observed~\cite{Coddington:2003}. Similarly to our toy model, there is one gapless excitation with a quadratic dispersion relation and one gapped mode with the cyclotron gap, $qB/m=2\Omega$~\cite{Sonin:1987,*Baym:2003}. The discussion in Section~\ref{subsec:discussion} suggests there is still an independent Bogoliubov mode stemming from spontaneous breaking of the U$(1)$ symmetry. We believe that our general arguments based solely on symmetry properties may lead to a new understanding of NG modes in the vortex lattice. 


\section{Summary and conclusions}
\label{sec:summary}

In the present paper, we have developed a framework for the analysis of spontaneous symmetry breaking in the case that there is only a discrete translational invariance. While our prime motivation was to study spontaneous breaking of continuous translational invariance, our results apply equally well to systems where this is broken explicitly by construction.

We generalized existing theorems on NG bosons in quantum many-body systems, thus completing the program initiated in our recent paper~\cite{WatanabeBrauner:2011}. Our main message is that as long as SSB is concerned, the usual division of symmetries into internal and spacetime ones is artificial. Instead, we propose that all symmetries that are uniform in the sense defined in Section~\ref{sec:general} can, and should, be treated on the same footing. This naturally includes (uniform) internal symmetries as well as translational invariance.

To illustrate our general arguments, we analyzed a model for a (ferromagnetic) supersolid in two spatial dimensions using the mean-field approximation at zero temperature. Using numerical computations, we examined the regime of weak coupling where the model features a homogeneous superfluid state, as well as the strong-coupling one where a supersolid crystalline order emerges. The properties of the spectrum of NG modes agree with our general conclusions in both phases. In particular, we have demonstrated that the Bogoliubov mode and the longitudinal phonon in the supersolid state are distinct physical modes.

Finally, we discussed the effect of an external magnetic field on the translational properties of the system. Using a simple quantum-mechanical toy model, we argued that this changes the spectrum of NG modes qualitatively by making the dispersion relation of some of them quadratic at low momentum, and giving a gap to others. This issue would certainly deserve a more detailed investigation in a quantum-field-theoretic setting, with respect to its intrinsic interest as well as the potential application to the physics of rotating Bose gases. 


\begin{acknowledgments}
We are grateful to H.~Abuki, Y.~Akamatsu, G.~Anagama, H.~Aoki, S.~Endo, Y.~Kato, Y.~Kawaguchi, M.~Kunimi, D.-H.~Lee, and E.~B.~Sonin for useful discussions. The work of T.B.~was supported by the Sofja Kovalevskaja program of the Alexander von Humboldt Foundation.
\end{acknowledgments}


\appendix

\section{Derivation of Eq.~\eqref{eq:S}}
\label{app:derivation}

The derivation of Eq.~\eqref{eq:S} is straightforward, if somewhat tedious. It bases solely on the partition of unity~\eqref{eq:partition}, translational property~\eqref{eq:uniform}, definition of the averaged charge density in Eq.~\eqref{eq:ucaverage}, and some elementary integrals. We would also like to first remind the reader of the Poisson identity for the periodic $\delta$-function in $\mathbb R^{d'}$,
\begin{equation}
\frac{\Omega_{\text{uc}}}{(2\pi)^{d'}}\sum_{\bm R}e^{i\bm k_1\cdot\bm R}=\sum_{\bm G}\delta^{d'}(\bm k_1-\bm G),
\label{eq:poisson}
\end{equation}
where $\bm G$ are vectors from the reciprocal lattice, $\bm G=\sum_{i=1}^{d'}n_i\bm b_i$ ($n_i\in\mathbb Z$ and $\bm a_i\cdot\bm b_j=2\pi\delta_{ij}$). Taking now the first term in the commutator in Eq.~\eqref{eq:S}, we obtain by a series of manipulations,
\begin{widetext}
\begin{equation}
\begin{split}
&\int d^{d+1}x\,\langle 0|j^0_a(x)\phi_b(0)|0\rangle\,e^{i(\omega t-\bm k\cdot\bm x)}\\
&=\int dt\int d^{d'}\bm{x}_1\int d^{d-d'}\bm{x}_2\sum_{\xi}\int_{(\mathrm{FBZ})}\frac{d^d\bm{q}}{(2\pi)^d}\langle 0|j^0_a(t,\bm{x}_1,\bm{x}_2)|\xi,\bm{q}\rangle\langle \xi,\bm{q}|\phi_b(0)|0\rangle\,e^{i(\omega t-\bm k\cdot\bm x)}\\
&=\int dt\sum_{\bm{R}}\int_{\mathrm{uc}}d^{d'}\bar{\bm{x}}_1\sum_{\xi}\int_{(\mathrm{FBZ})}\frac{d^d\bm{q}}{(2\pi)^d}\langle 0|j^0_a(t,\bar{\bm{x}}_1+\bm{R},\bm{0})|\xi,\bm{q}\rangle\langle \xi,\bm{q}|\phi_b(0)|0\rangle\,e^{i(\omega t-\bm{k}_1\cdot\bm{x}_1)}\left[\int d^{d-d'}\bm{x}_2\,e^{i(\bm{q}_2-\bm{k}_2)\cdot\bm{x}_2}\right]\\
&=\int dt\sum_{\bm{R}}\int_{\mathrm{uc}}d^{d'}\bar{\bm{x}}_1\sum_{\xi}\int_{\mathrm{FBZ}}\frac{d^{d'}\bm{q}_1}{(2\pi)^{d'}}\langle 0|j^0_a(0,\bar{\bm{x}}_1,\bm{0})|\xi,\bm{q}'\rangle\langle \xi,\bm{q}'|\phi_b(0)|0\rangle\,e^{i(\bm{q}_1-\bm{k}_1)\cdot\bm{R}-i\bm{k}_1\cdot\bar{\bm{x}}_1-i[\varepsilon_{\xi}(\bm{q}')-\omega]t}\\
&=\sum_{\xi}\int_{\mathrm{FBZ}}d^{d'}\bm{q}_1\,\langle 0|\left[\int_{\mathrm{uc}}\frac{d^{d'}\bar{\bm{x}}_1}{\Omega_{\mathrm{uc}}}j^0_a(0,\bar{\bm{x}}_1,\bm{0})e^{-i\bm{k}_1\cdot\bar{\bm{x}}_1}\right]|\xi,\bm{q}'\rangle\langle \xi,\bm{q}'|\phi_b(0)|0\rangle\,2\pi\delta(\omega-\varepsilon_{\xi}(\bm{q}'))\left[\dfrac{\Omega_{\mathrm{uc}}}{(2\pi)^{d'}}\sum_{\bm{R}}e^{i(\bm{q}_1-\bm{k}_1)\cdot\bm{R}}\right]\\
&=\sum_{\xi}\int_{\mathrm{FBZ}}d^{d'}\bm{q}_1\,\langle 0|\overline{j_a}_{(\bm{k})}|\xi,\bm{q}'\rangle \langle \xi,\bm{q}'|\phi_b(0)|0\rangle\,2\pi\delta(\omega-\varepsilon_{\xi}(\bm{q}'))\sum_{\bm{G}}\delta^{d'}(\bm{q}_1-\bm{k}_1-\bm{G})\\
&=2\pi\sum_{\xi}\langle 0|\overline{j_a}_{(\bm{k})}|\xi,\bm{k}\rangle\langle \xi,\bm{k}|\phi_b(0)|0\rangle\delta(\omega-\varepsilon_{\xi}(\bm{k})),
\end{split}
\end{equation}
\end{widetext}
where we denoted $\bm{q}'=(\bm{q}_1,\bm{k}_2)$. In the last step, we used the fact that, similarly to the coordinate decomposition $\bm x_1=\bar{\bm x}_1+\bm R$, any vector $\bm k_1$ from the reciprocal space can be uniquely expressed as a sum of a vector from FBZ and a vector $\bm G$ from the reciprocal lattice. As long as $\bm k_1$ already lies in FBZ, the only solution to the condition $\bm{q}_1-\bm{k}_1-\bm{G}=\bm0$ is $\bm q_1=\bm k_1$ and $\bm G=\bm0$. The second contribution to the commutator in Eq.~\eqref{eq:S}, $\int d^{d+1}x\,\langle 0|\phi_b(0) j^0_a(x)|0\rangle\,e^{i(\omega t-\bm k\cdot\bm x)}$, can be calculated in the same way and we therefore skip the details.


\section{Goldstone theorem in the effective action formalism}
\label{app:effaction}

In Appendix~B of the paper~\cite{WatanabeBrauner:2011} we provided a proof of the Goldstone theorem based on the quantum effective action in a form applicable to nonuniform symmetries. However, we for the sake of simplicity assumed continuous translational invariance of the ground state. Our aim here is to relax this constraint. Like in the rest of this paper, we will assume that the spontaneously broken symmetry is uniform; the extension of our argument to the fully general case is straightforward though.

Consider a theory of a set of (not necessarily scalar) fields, $\phi_i(x)$, whose classical action is invariant under the infinitesimal transformation $\delta\phi_i(x)=\theta F_i[\phi(x)]$. Here $\theta$ is a parameter of the transformation and $F_i$ is a local functional of the fields that does not depend explicitly on the coordinate $x$. (This is equivalent to the requirement that the symmetry be uniform.) Provided that $F_i$ is linear in the fields, the quantum effective action, $\Gamma[\phi]$, shares the symmetry of the classical theory~\cite{Weinberg:1996v2}. We then obtain the invariance condition
\begin{equation}
\int d^{d+1}y\,\frac{\delta^2\Gamma[\phi_0]}{\delta\phi_i(x)\delta\phi_j(y)}F_j[\phi_0(\bm y)]=0,
\label{eq:flatdir}
\end{equation}
where the integration measure involves time as well as $d$ spatial coordinates and $\phi_{0i}(\bm{x})$ denotes the vacuum expectation value of the field in accord with the notation introduced in Section~\ref{subsec:classical}. The second functional derivative of the effective action here represents the exact inverse propagator of the theory, $\mathcal G^{-1}_{ij}(x,y)$. Assuming as in the rest of the paper a discrete translational invariance in $\mathbb R^{d'}$ and continuous one in $\mathbb R^{d-d'}$ as well as in time, this can be Fourier transformed as
\begin{equation}
\begin{split}
\mathcal G^{-1}_{ij}&(x,y)=\int_{-\infty}^{+\infty}\frac{d\omega}{2\pi}\int_{\text{(FBZ)}}\frac{d^d\bm{k}}{(2\pi)^d}\,e^{-i\omega(t_x-t_y)}\\
&\times e^{i[\bm k_1\cdot(\bm R_x-\bm R_y)+\bm k_2\cdot(\bm x_2-\bm y_2)]}\mathcal G^{-1}_{ij}(\omega,\bm k;\bar{\bm x}_1,\bar{\bm y}_1),
\label{eq:FT}
\end{split}
\end{equation}
where $\bm k_1$ lies in FBZ. Note that the Fourier transform of the propagator still depends on the coordinates within the Wigner--Seitz unit cell as a result of the lack of full continuous translational invariance. Thanks to the assumed periodicity of the ground state, $\phi_{0i}(\bm y)=\phi_{0i}(\bar{\bm y}_1)$, Eq.~\eqref{eq:flatdir} becomes
\begin{equation}
\int_{\text{uc}}\frac{d^{d'}\bar{\bm y}_1}{\Omega_{\text{uc}}}\,\mathcal G^{-1}_{ij}(0,\bm0;\bar{\bm x}_1,\bar{\bm y}_1)F_j[\phi_0(\bar{\bm y}_1)]=0.
\label{eq:invprop}
\end{equation}
For a translationally invariant ground state, $F_j[\phi_0(\bar{\bm y}_1)]$ is a constant and it follows immediately that the inverse propagator must have a gapless pole at zero momentum once the symmetry is spontaneously broken, that is, $F_j[\phi_0]$ is nonzero. However, in presence of a spatially varying order parameter, the implications of Eq.~\eqref{eq:invprop} for the excitation spectrum are less straightforward.

It is useful to first recall the spectral representation of the propagator. Using the definition of the propagator in the operator formalism, $i\mathcal G_{ij}(x,y)=\langle0|T\{\phi_i(x)\phi_j(y)\}|0\rangle$, and inserting the partition of unity~\eqref{eq:partition}, one arrives at the following expression for the Fourier transformed propagator, defined analogously to Eq.~\eqref{eq:FT},
\begin{equation}
\begin{split}
\mathcal G_{ij}&(\omega,\bm k;\bar{\bm x}_1,\bar{\bm y}_1)=\sum_\xi\\
\times\biggl[&\frac{\langle0|\phi_i(0,\bar{\bm x}_1,\bm0)|\xi,\bm k\rangle\langle\xi,\bm k|\phi_j(0,\bar{\bm y}_1,\bm0)|0\rangle}{\omega-\frac1\hbar\varepsilon_\xi(\bm k)+i0}\\
&-\frac{\langle0|\phi_j(0,\bar{\bm y}_1,\bm0)|\xi,-\bm k\rangle\langle\xi,-\bm k|\phi_i(0,\bar{\bm x}_1,\bm0)|0\rangle}{\omega+\frac1\hbar\varepsilon_\xi(-\bm k)-i0}\biggr].
\end{split}
\label{eq:kallen}
\end{equation}
This ensures that even with just a discrete translational invariance, one can still extract the quasiparticle spectrum from the poles of a conveniently defined propagator. To conclude our argument, we note that the condition of $\mathcal G^{-1}$ being inverse to $\mathcal G$ reads $\int d^{d+1}z\,\mathcal G_{ik}(x,z)\,\mathcal G_{kj}^{-1}(z,y)=\delta_{ij}\delta^{d+1}(x-y)$, which takes the following form in the frequency--momentum space,
\begin{equation}
\begin{split}
\int_{\text{uc}}\frac{d^{d'}\bar{\bm z}_1}{\Omega_{\text{uc}}}\,\mathcal G_{ik}(\omega,\bm k;\bar{\bm x}_1,\bar{\bm z}_1)\,\mathcal G^{-1}_{kj}(\omega,\bm k;\bar{\bm z}_1,\bar{\bm y}_1)\\
=\delta_{ij}\Omega_{\text{uc}}\delta^{d'}(\bar{\bm x}_1-\bar{\bm y}_1).
\end{split}
\label{eq:last}
\end{equation}
Multiplying this equation by $F_j[\phi_0(\bar{\bm y}_1)]$ and integrating over $\bar{\bm y}_1$, we assert with the help of Eq.~\eqref{eq:invprop} and the spectral representation~\eqref{eq:kallen} that as soon as $F_j[\phi_0(\bar{\bm x}_1)]$ is nonzero at some $\bar{\bm x}_1$, the propagator (at the same point) must have a pole such that $\varepsilon_\xi(\bm0)=0$.

Finally, let us remark that the proof would have been technically much simpler had we considered the effective action (and in turn the propagator) as a functional of fields spatially averaged over the Wigner--Seitz unit cell,
\begin{equation}
\overline\phi_i(t,\bm R,\bm x_2)\equiv\int_{\text{uc}}\frac{d^{d'}\bar{\bm x}_1}{\Omega_{\text{uc}}}\,\phi_i(t,\bar{\bm x}_1+\bm R,\bm x_2).
\end{equation}
The order parameter $\overline\phi_{0i}$ would then be completely time and coordinate independent and the left-hand sides of Eqs.~\eqref{eq:invprop} and~\eqref{eq:last} would become mere products in momentum space. The price for this simplification would be that we could make no conclusions about the spectrum in the cases that the symmetry is spontaneously broken but the spatial average of the order parameter vanishes. The above given proof is more general since it only assumes that the order parameter is nonzero at some point.


\section{Calculational methods}
\label{app:methods}

In this appendix, we collect some details of the techniques used in the numerical computations in Section~\ref{sec:supersolid}.


\subsection{Minimization of the energy}
\label{app:minimize}

To minimize the energy functional~\eqref{eq:E}, we exploit the assumed periodicity of the classical field and perform the Fourier decomposition of the field, $\psi_{\bm{G}}=\Omega_{\mathrm{uc}}^{-1}\int_{\mathrm{uc}}d^d\bm{x}\,\psi(\bm{x})e^{-i\bm{G}\cdot\bm{x}}$. The total energy divided by the space volume is then equal to the energy density averaged over the unit cell, which is given by
\begin{equation}
\begin{split}
\bar{\mathcal{E}}[\psi]=&\sum_{\bm{G}}\left(\frac{\bm G^2}{2g}-1\right)\psi_{i\bm{G}}^*\psi_{i\bm{G}}\\
&+\frac{1}{2}\sum_{\bm{G},\bm{G}'\bm{q}}\psi_{i\bm{G}}^*\psi_{i\bm{G}-\bm{q}}v(\bm{q})\psi_{j\bm{G}'}^*\psi_{j\bm{G}'+\bm{q}}.
\end{split}
\end{equation}
Thanks to the parity invariance of the Lagrangian, which is naturally assumed to be shared by the ground state, we can restrict our attention to even-parity classical fields, $\psi_0(\bm{x})=\psi_0(-\bm{x})$. This, together with the reality of the classical field, implies the relation $\psi_{\bm{G}}=\psi_{-\bm{G}}=\psi_{\bm{G}}^*$, which significantly reduces the number of variables in the minimization procedure.

The minimization of the averaged energy density is accomplished by first fixing the primitive lattice vectors $\{\bm{a}_i\}_{i=1}^d$ and minimizing $\bar{\mathcal{E}}[\psi]$ with respect to the Fourier components $\{\psi_{\bm{G}}\}$. This determines the function $\bar{\mathcal{E}}(\{\bm{a}_i\})\equiv \bar{\mathcal{E}}[\psi_0(\{\bm{a}_i\})]$, which is subsequently minimized with respect to $\{\bm{a}_i\}$. The actual crystal form of the classical field $\psi_0(\bm{x})$ is thus found.


\subsection{Calculation of the band structures}
\label{app:band}

The Schr\"odinger equation~\eqref{eq:Sch} can be rewritten in Fourier space as
\begin{equation}
\begin{split}
T_{\bm{G}_1\bm{G}_2}^*\varphi_{i\bm{G}_2}&=-(K_{\bm{G}_1\bm{G}_2}+2\delta_{i1}W_{\bm{G}_1\bm{G}_2})\phi_{i\bm{G}_2},\\
T_{\bm{G}_1\bm{G}_2}\phi_{i\bm{G}_2}&=-K_{\bm{G}_1\bm{G}_2}\varphi_{i\bm{G}_2},
\end{split}
\label{eq:Sch2}
\end{equation}
where the (infinite-dimensional) matrices $K$, $T$ and $W$ are defined by 
\begin{equation}
\begin{split}
K_{\bm{G}_1\bm{G}_2}&=\left[\frac{(\bm{k}+\bm{G}_1)^2}{2g}-1\right]\delta_{\bm{G}_1\bm{G}_2}+u_{\bm{G}_1-\bm{G}_2},\\
T_{\bm{G}_1\bm{G}_2}&=i\omega(\bm{k})\delta_{\bm{G}_1\bm{G}_2},\\
W_{\bm{G}_1\bm{G}_2}&=\sum_{\bm{G}'}\bar\psi_{0\bm{G}_1-\bm{G}'}\bar\psi_{0\bm{G}'-\bm{G}_2}v(\bm{k}+\bm{G}'),
\end{split}
\end{equation}
and $u_{\bm{G}}=v(\bm{G})\sum_{\bm{G}'}\bar\psi_{0\bm{G}-\bm{G}'}\bar\psi_{0\bm{G}'}$. Because Eq.~\eqref{eq:Sch2} represents a set of homogeneous linear equations for the Fourier components $\phi_{i\bm G}$ and $\varphi_{i\bm G}$ (for each fixed $i$), the band structure can be found from the condition that the determinant of the matrix of coefficients in these equations be zero.


\bibliography{references}

\end{document}